\documentclass[oneside,reqno,12pt]{article}
\usepackage{amsmath}

\title{
 Eigenvalue Repulsion\\
 and Matrix Black Holes
}
\author{
 Gavin Polhemus
 \thanks{\texttt{g-polhemus@uchicago.edu}}
}
\date{
 \textit{
  Department of Physics and Enrico Fermi Institute\\
  The University of Chicago\\
  5640 Ellis Avenue, Chicago, Illinois 60637
 }\\
}

\newcommand{\EFInumber}{EFI-99-14}
\newcommand{\archivenumber}{hep-th/9904128}

\newcommand{\ie}{i.e.\ }
\newcommand{\infinity}{\infty}

\newcommand{\Tr}{\mathop{\rm Tr}\nolimits}

\newcommand{\hc}[1]{{{#1}^\dag}}

\newcommand{\lp}{l_{\text{p}}}

\newcommand{\Gn}{G_{N}}
\newcommand{\Rs}{R_{s}}

\newcommand{\greekdef}[3]
 {\def#1{\relax\ifmmode#2\else#3\fi}}

\greekdef{\o}{\theta}{\char"1C}
\greekdef{\i}{\iota}{\char"10}

\greekdef{\l}{\lambda}{\char32l}

\newcommand{\D}{\Delta}
\greekdef{\O}{\Theta}{\char"1F}
\greekdef{\L}{\Lambda}
 {\leavevmode\setbox0\hbox{L}\hbox to\wd0{\hss\char32L}}

\greekdef{\P}{\Pi}{\mathhexbox27B}
\greekdef{\S}{\Sigma}{\mathhexbox278}

\begin{document}
 
\setlength{\unitlength}{1mm}

%\section{Title Page}
\begin{titlepage}

\maketitle
\thispagestyle{empty}

\begin{table}[t!]
\rightline{\archivenumber}
\rightline{\EFInumber}
\end{table}

\begin{abstract}
 Eigenvalue repulsion can explain the holographic growth of black 
 holes in Matrix theory.
 The resulting picture is essentially the same as the 
 Boltzman gas picture but avoids any assumption about the effective 
 potential between the D0 branes.
 Further, eigenvalue 
 repulsion extends the Boltzman gas picture past the 
 BFKS point to $N\gg S$.
	The use of Boltzman statistics is natural in this picture.
\end{abstract}

\end{titlepage}
\setcounter{footnote}{0}

\section{Introduction}

Since Matrix theory is conjectured to be the discrete light-cone 
quantization of M-theory, it is natural to use it as a way of 
investigating the quantum properties of black holes.

Banks, Fischler, Klebanov and Susskind (BFKS)
and Martinec and Li, have studied black holes whose entropy, $S$, roughly 
the same as their light-like momentum, $N$
\cite{Banks:1998hz,Klebanov:1997kv,Banks:1998tn}.
They can be formed by reducing the energy of a highly excited cluster 
of D0 branes until the momentum of the individual D0 branes is the 
inverse of the cluster's transverse size (saturating the Heisenberg uncertainty 
bound).
The transverse size is found by setting the potential equal to the kinetic 
energy (in accord with the virial theorem).  
The choice of potential is crucial.  The obvious 
choice, $v^{4}\!/r^{7}$, gives the Schwarzschild radius, $\Rs$.
However, at this point the expansion parameter 
for the potential is 1, so the expansion is on the verge of breaking down
\cite{Horowitz:1998fr}.

Nonetheless, once the correct radius is found, Matrix theory makes 
many correct predictions about black holes:
the right mass for a given radius 
\cite{Klebanov:1997kv,Horowitz:1998fr,Banks:1998tn}
and the correct long range gravitational 
potential between equal mass black holes \cite{Banks:1998tn}.
The constituents have the correct properties to be Hawking 
radiation \cite{Banks:1998tn}.
If the particles are treated with Boltzman statistics, 
the entropy is even correct
\cite{Volovich:1996gw,Horowitz:1998fr,Banks:1998tn}. 

The problem becomes more pronounced when $N \gg S$.  Simply 
replacing the D0 branes with bound states and using the $v^{4}/r^{7}$
potential fails to give correct results.
Li identified other terms in the matix theory effective potential that 
could give the correct  radius, but it is not clear why those terms would 
dominate \cite{Li:1998iz}.
Li and Martinec get the correct results using $v^{4}/r^{7}$ to 
approximate processes that exchange longitudinal momentum
\cite{Li:1998ci}.

Eigenvalue repulsion can be used to find the size of black holes 
without any knowledge of the potential.
Using this size, all of the previous black hole results follow
\cite{Horowitz:1998fr,Li:1998ci,Gao:1998fe}.
Eigenvalue repulsion allows the results to be extended to $N \gg S$.
Because eigenvalue repulsion involves the off-diagonal elements of 
the matrices, it will be natural to treat the system with Boltzman 
statistics.

% This paper is not rigorous and many ideas are probably greatly 
% simplified.  However, this sketch contains many important 
% elements of matrix black holes.
First, I describe eigenvalue repulsion in Matrix theory.
Then I find the radius and energy at $R=\Rs$ and discuss entropy in this 
case.
Finally, I discuss $R \gg \Rs$.
All factors of order one are ignored throughout.

\section{Low energies and eigenvalue repulsion}

To study black holes we need to understand matrix theory in the 
region where the energy per D0 brane is very low (much less than the 
plank energy).  This will mean two things:
First, remote D0 branes are not connected by strings 
(\ie the off-diagonal elements between well separated D0 branes are in the 
ground state).

As the D0 branes get within $\lp$ of each other, the quantum fluctuations in 
the off diagonal elements connecting the D0 branes also become of 
order $\lp$, so the D0 branes are able to explore their full matrix 
degrees of freedom.
This region is sometimes called the stadium.
The second consequence of the low energy is that the wave-function must be 
almost constant across the width of the stadium since variations on 
this scale would require very large kinetic energies.

The flatness of the wave function in the stadium means that we can 
look to the theory of random matrices for intuition about how the 
system will explore the full matrix degrees of freedom.
It is known that the statistics of random matrices strongly favors matrices 
with well separated eigenvalues, pushing the D0 branes to the edge of 
the stadium.

The role of eigenvalue repulsion can be made a bit more quantitative.
In the stadium, the distance between two D0 branes is $r^{2} = -\Tr X^{i}X^{i}$.
For fixed $r$, this is the equation for a sphere in 27 dimensions 
(nine directions times the three independent generators of traceless, 
antihermitian, $2 \times 2$ matrices).
Since the wave function is relatively flat in this region, the probability 
that the D0 branes are a distance $r$ apart falls off like $r^{26}$.
That is 18 powers of $r$ faster than the $r^{8}$ probability that one
would expect for the nine spatial directions.
This means that the wave function is strongly dominated by 
configurations which have the D0 branes near the edge of the stadium.
There is no actual potential pushing the D0-branes apart, but this 
statistical effect is the Matrix Theory manifestation of eigenvalue 
repulsion.%
\footnote{
	The gauge symmetry does not affect this result.
 The gauge field degrees of freedom are removed by going to $A=0$ gauge.
 The assumption that the wave function is flat satisfies the 
 requirement that the state be annihilated by the gauge generators.
}

The situation is similar when $N$ D0 branes are present.
When the separation between \emph{any two} D0 branes falls below $\lp$, new 
matrix degrees of freedom open up and eigenvalue repulsion turns on.
This causes the wave function to strongly favor configurations with size at 
least $N^{1/9}\lp$, one D0 brane per plank volume.

It has been shown that composite systems of D0 branes grow like 
$N^{1/3}$, much faster than the $N^{1/9}$ growth predicted by 
eigenvalue repulsion
\cite{Susskind:1998vk}.
However, this rapid growth is caused by fluctuations in very high energy, 
off-diagonal elements of the matrix connecting widely separated D0 
branes.
Only very high energy probes will be able to see these high frequency 
quantum fluctuations.
Eigenvalue repulsion is a zero energy effect, so it should be seen by any 
probe.
Similarly, a low energy probe will be captured by a black hole if its 
impact parameter is less than $R_{s}$, but a high energy probe (with 
energy much greater than the mass of the black hole) will be captured 
at a much greater distance.

\section{Black holes at the BFKS point}

A very highly excited clump of D0 branes has enough mass to create a 
black hole whose radius is larger than the light-like compactification 
radius.
Lowering the system's energy will slow the constituent D0 branes until 
the momentum saturates the Heisenberg uncertainty bound, \ie the 
momentum of individual D0 branes is one over the size of the clump.
This is the BFKS point
\cite{Banks:1998hz}.
Using the size, $N^{1/9}\lp$, given by eigenvalue 
repulsion one can find the mean kinetic energy. 
\begin{align}
 E_{k}
 &= \frac{R}{\hbar} \Tr \P^{i}\P^{i}
 \\
 &\approx \frac{R}{\hbar} \sum_{a=1}^{N} \frac{\hbar^{2}}{\Rs^{2}}
 \\
 &\approx \frac{R \Rs^{7}}{\Gn}
\end{align}

We know from the virial theorem that the potential energy will be of 
the same order as the kinetic energy, so we do not need to know anything 
about its form.

Other than the form of the potential, this is exactly the system 
studied by previous authors.
As they have observed, this system has the right relationship between mass 
and radius
\cite{Klebanov:1997kv,Horowitz:1998fr,Banks:1998tn},
\begin{equation}
 M^{2} = E\frac{N\hbar}{R} = \frac{\Rs^{16}}{\Gn^{2}},
\end{equation}
and gives the correct long range gravitational interaction for 
equal mass black holes \cite{Banks:1998tn}.
The constituent D0 branes have the right energy and momentum to be Hawking 
radiation, should they escape \cite{Banks:1998tn}.

\section{Entropy}

To get the correct entropy the D0 branes must be treated with 
Boltzman statistics
\cite{Banks:1998tn}.
This occurs because the wave function is dominated by states that 
break all of the gauge symmetry mixing remote D0 branes,
even the permutation symmetry.

The large separation between the majority of D0 branes breaks the continuous 
symmetry down to the permutation symmetry.  If all the D0 branes were 
far apart one could gauge transform all of the matrices into diagonal 
form simultaneously.  Since this diagonalization would only be unique 
up to permutations, the permutation symmetries would be unbroken.

% Eigenvalue repulsion causes localization without the usual large 
% kinetic energy
% This breaks the statistics symmetry.
However, since neighboring D0 branes are near enough to explore the 
full matrix degrees of freedom, the off 
diagonal elements connecting them cannot be integrated out
\cite{Li:1998ci}.
The matrices cannot be simultaneously diagonalized, only put in band 
diagonal form.%
\footnote{
 Here I am using ``band diagonal'' in a very general sense.
	The D0 branes will be connected to their nearest neighbors in all nine 
	dimensions.  There will be no way to write these matrices as two 
	dimensional arrays with zeros away from the diagonal.
}
This breaks the permutation symmetry because a gauge transformation 
that permutes 
remote D0 branes would take the matrices out of band diagonal form.

For example, consider a situation where D0 branes $a,b$ and $c$ are 
close together and $e,f$ and $g$ are close together, but the two sets 
are far apart.  This will be represented by matrices like this 
(All the 1's represent terms of order $\lp$, not literally 1):
\begin{equation}
	X^i = 
 \begin{bmatrix}
  \ddots & 1 &   &   &      &        &   &   &   &    &\\
	 1    & x^i_a & 1 &   &      &        &   &   &   &    &\\
       & 1 & x^i_b & 1 &      &        &   &   &   &    &\\
       &   & 1 & x^i_c & 1    &        &   &   &   &    &\\
      &   &   & 1  & \ddots &	\ddots &   &   &   &    &\\
     &   &   &   & \ddots &	\ddots & 1  &   &   &     &\\
	    &   &   &	  &        & 1    & x^i_e & 1 &   &      &\\
	    &   &   &	  &        &      & 1 & x^i_f & 1 &      &\\
	    &   &   &	  &        &	     &   & 1 & x^i_g &    1 &\\
					&   &   &	  &        &	     &   &   & 1 & \ddots &\\
\end{bmatrix}
 \end{equation}

Permuting the $b$ and $f$ D0 branes takes the matrices out of band diagonal form:
 \begin{equation}
		\hc{U} X^i U =
		\begin{bmatrix}
 		\ddots & 1   &   &   &         &        &   &   &   &    &\\
		 1    & x^i_a &   &   &         &        &   & 1 &   &    &\\
 	      &     & x^i_f &   &       &        & 1 &   & 1 &    &\\
		     &     &      & x^i_c & 1   &        &   & 1 &   &    &\\
	      &    &     & 1    & \ddots &	\ddots &   &   &   &    &\\
		    &    &    &     & \ddots &	\ddots & 1  &   &   &     &\\
	     &   & 1 &	   &        & 1    & x^i_e &   &   &      &\\
	     & 1 &   &	1 &        &      &   & x^i_b &   &      &\\
 	    &   & 1 &	  &        &	     &   &   & x^i_g &    1 &\\
		    &   &   &	  &        &	     &   &   & 1 & \ddots &\\
 	\end{bmatrix}
 \end{equation}
where $U$ is the matrix permuting $b$ and $f$.

The symmetries mixing neighboring D0 branes is not broken at all. 
However, the number of near neighbors is of order one, so in doing 
statistical machanics for large $N$ the D0 branes should be treated as 
distinguishable.

This description is similar to the proposal of BFKS  
in which the D0 branes are tethered to a background configuration
\cite{Banks:1998tn}.
Here, however, there is no firm distinction between the background and 
the excitations.

\section{Beyond the BFKS point}

At the BFKS point the constituent D0 branes have already saturated their 
uncertainty bound, so the energy of the system cannot be lowered by 
further reducing their momentum.
They will have to band together into bound states.
These bound states will have a larger mass (in the Matrix theory, 
they are of course massless in M-theory) and will be able to live in 
a smaller volume with less energy \cite{Horowitz:1998fr}.

The arguments of the previous section would appear to apply to the bound 
states of Matrix theory, causing them to be seen by low energy probes as 
large objects.
However, even in the large $N$ limit, these objects should 
not have hard interactions for impact parameters greater than $\lp$.
Some miracle of the bound state wave function and supersymmetry must cause 
cancellations that allow the objects to pass through each other without 
interaction.

These miracles will not occur when the wave functions are excited.  
In eleven dimensions this corresponds to changing the supergravitons 
into black holes.
Excitations will cause the graviton to break into a metastable 
collection of $n$ smaller gravitons.
Each of these is unexcited, and is therefore still able to 
pass through the others without interaction.  The distance at which 
the eigenvalue repulsion plays a role is still $\lp$.

Consider a system of two well separated bound states, consisting of 
$m_{1}$ and $m_{1}$ D0 branes each.  They will break the 
$U(m_{1}+m_{2})$ symmetry down to $U(m_{1}) \times U(m_{2})$.  There 
are $m_{1}m_{2}$ strings connecting them, each with 8 transverse, 
complex polarizations.  The remaining $U(m_{1}) \times U(m_{2})$ gauge 
symmetry and the $O(8)$ rotation symmetry about the axis of separation 
is enough to relate all of these off diagonal modes, so they must all 
have the same mass.  The mass is given by the separation of the bound 
states exactly as in the case of two separated D0 branes.  When the 
centers of the two bound states get within a distance $\lp$ of each 
other, all $16 m_{1}m_{2}$ of these degrees of freedom open up, causing 
tremendous eigenvalue repulsion.

This bound state cluster picture closely mimics the description at the BFKS 
point, and has been studied by several previous authors
(once the radius is found).
Once the radius is found, many of the successes of $N = S$ can be carried 
over in a straightforward  manner to $N \gg S$.
In particular, the mass to radius relationship is 
correct, the constituent bound states 
have the right properties to become Hawking radiation, and 
Boltzman statistics give the correct entropy \cite{Li:1998ci}.

The breaking of the statistics symmetries by off diagonal 
fluctuations should work for bound states exactly as it did for 
individual D0 branes, explaining the use of Boltzman statistics.

\subsection{Newtonian Potential}

As a demonstration, consider the potential between two static black 
holes of different masses.%
\footnote{
 This analysis is similar to the equal mass case studied by Banks 
 et.~al.~\cite{Banks:1998tn},	but differs from
 Gao and Zhang's treatment \cite{Gao:1998fe}.
}
We will assume that their separation is 
greater than $R$, where $v^{4}/r^{7}$ dominates the two body 
interactions.

It will be much more convenient to use the velocities of the bound 
states rather than their momenta.  The mass of a bound state is 
$ \hbar m / R $.  Since the bound states that make up a black hole 
have momenta $ \hbar / \Rs $ the velocities are $ R / \Rs m $.  

First, observe that the velocity of the constituent bound states is the 
boost parameter required to bring the black hole to the rest frame.
\begin{align}
 \frac{E}{P_{-}}
 &=
 \frac{ER}{N\hbar}
 \\
 &= \frac{R^{2}\Rs^{7}}{\Gn m N \hbar}
 \\
 &= \frac{R^{2}}{m^{2} \Rs^{2}}
 \\
 &= v^{2}
\end{align}
Black holes that 
are at rest relative to each other will be made of bound states with 
the same velocity, though not of the same momentum.

It will be useful to note:
\begin{align}
	M^{2} &= E P_{-} = v^{2} P_{-}^{2} \\
	M &= \frac{v N \hbar}{R}
\end{align}

Next, we find the energy shift due to the $v^{4}/r^{7}$ interaction.  
Since the velocities of the constituents are roughly the same in the 
two black holes, $(v_{1}-v_{2})^{4}=v^{4}$.
\begin{align}
 \D E
 &=
 \frac{\Gn\hbar^{2}}{R^{3}}
 \sum_{n_{1},n_{2}} m_{1}m_{2}
 \frac{(v_{1}-v_{2})^{4}}{r^{7}}
 \\
 &= 
 \frac{\Gn\hbar^{2}}{R^{3}}
 N_{1}N_{2}
 \frac{v^{4}}{r^{7}}
 \\
 &= 
 \Gn \frac{M_{1}M_{2} v^{2}}{R r^{7}}
\end{align}

Finally, this can be used to find the potential.  The potential is 
much smaller than the masses of the black holes.
\begin{align}
 E\frac{N\hbar}{R}
 &=
 M^{2} = (M_{1} + M_{2} + V)^{2}
 \\
 V
 &=
 \frac{\D E (N_{1} + N_{2})\hbar}{R (M_{1} + M_{2})}
 \\
 &=
 \frac{\D E}{v}
 \\
 &= 
 \Gn \frac{M_{1}M_{2}}{R' r^{7}},
\end{align}
where $R'=R/v$ is the compactification radius in the rest frame of the 
black holes.

\section{Conclusions}

Eigenvalue repulsion predicts the size of black holes without requiring 
knowledge of the effective potential between the constituent D0 
branes, and suggests a mechanism for breaking the statistics symmetry 
of the constituents.  However, this understanding is still rather 
primitive.  Much work will have to be done even to turn the sketch 
presented here into a derivation of black hole thermodynamics.

Little has been said about the fermion degrees of freedom in this theory.
Might supersymmetry cause terms that cancel the eigenvalue repulsion?
This does not appear to happen.  
Supersymmetry requires that fermionic ground state energy cancels bosonic 
ground state energy. 
However, eigenvalue repulsion is purely 
statistical and is not affected by the fermionic states.

I would like to thank Jeff Harvey, Emil Martinec and Miao Li for 
helpful conversations about these matters.

\providecommand{\href}[2]{#2}\begingroup\raggedright\endgroup

% \bibliographystyle{utphys2}
% \bibliography{BlackEigen}

\end{document}